\documentclass[linenumbers, showpacs]{aastex631}

\usepackage{natbib}
\usepackage{booktabs}
\usepackage{lineno}

\begin{document}
\nolinenumbers
\baselineskip=0.8cm
\title{BCDDM: Branch-Corrected Denoising Diffusion Model for Black Hole Image Generation}


\author{Ao liu}
\altaffiliation{These authors contributed equally to this work.}
\affiliation{College of information science and engineering, Hunan Normal University, Changsha,410081, People's Republic of China}

\author{Zelin Zhang}
\altaffiliation{These authors contributed equally to this work.}

\affiliation{Department of Physics, Institute of Interdisciplinary Studies, Key Laboratory of Low Dimensional Quantum Structures and Quantum Control of Ministry of Education, Synergetic Innovation Center for Quantum Effects and Applications, Hunan Normal University,  Changsha, Hunan 410081, People's Republic of China}

\author{Songbai Chen}
\altaffiliation{Corresponding author: csb3752@hunnu.edu.cn}
\affiliation{Department of Physics, Institute of Interdisciplinary Studies, Key Laboratory of Low Dimensional Quantum Structures and Quantum Control of Ministry of Education, Synergetic Innovation Center for Quantum Effects and Applications, Hunan Normal University,  Changsha, Hunan 410081, People's Republic of China}
\affiliation{Center for Gravitation and Cosmology, College of Physical Science and Technology, Yangzhou University, Yangzhou 225009, People's Republic of China}

\author{Cuihong Wen}
\altaffiliation{Corresponding author:cuihongwen@hunnu.edu.cn}
\affiliation{College of information science and engineering, Hunan Normal University, Changsha,410081, People's Republic of China}

\author{Jieci Wang}
\altaffiliation{Corresponding author:jieciwang@hunnu.edu.cn}
\affiliation{Department of Physics, Institute of Interdisciplinary Studies, Key Laboratory of Low Dimensional Quantum Structures and Quantum Control of Ministry of Education, Synergetic Innovation Center for Quantum Effects and Applications, Hunan Normal University,  Changsha, Hunan 410081, People's Republic of China}


\begin{abstract}
\nolinenumbers
    \baselineskip=0.6 cm
    The properties of black holes and accretion flows can be inferred by fitting Event Horizon Telescope (EHT) data to simulated images generated through general relativistic ray tracing (GRRT). However, due to the computationally intensive nature of GRRT, the efficiency of generating specific radiation flux images needs to be improved. This paper introduces the Branch Correction Denoising Diffusion Model (BCDDM), a deep learning framework that synthesizes black hole images directly from physical parameters. The model incorporates a branch correction mechanism and a weighted mixed loss function to enhance accuracy and stability. We have constructed a dataset of 2,157 GRRT-simulated images for training the BCDDM, which spans seven key physical parameters of the radiatively inefficient accretion flow (RIAF) model. Our experiments show a strong correlation between the generated images and their physical parameters. By enhancing the GRRT dataset with BCDDM-generated images and using ResNet50 for parameter regression, we achieve significant improvements in parameter prediction performance. BCDDM offers a novel approach to reducing the computational costs of black hole image generation, providing a faster and more efficient pathway for dataset augmentation, parameter estimation, and model fitting.
\end{abstract}

\newpage

\section{Introduction}
The recent release of black hole images by the Event Horizon Telescope (EHT) collaboration provides direct evidence for the existence of black holes in the universe \citep{2019ApJ...875L...1E,2022ApJ...930L..12E,2024A&A...681A..79E}. These observations offer a unique opportunity to probe the electromagnetic interactions, material distribution, and accretion processes near black holes, further testing the validity of gravity theories and accretion disk models \citep{2018NatAs...2..585M,2021ApJ...910L..12E,2021ApJ...910L..13E,2021ApJ...912...35N,2024ApJ...964L..26E,2024ApJ...964L..25E,2022ApJ...930L..17E,2022ApJ...930L..16E,2022SCPMA..6520411L,2022CoTPh..74i7401W,2022EPJC...82..835Z,2022ApJ...938....2Q,2023SCPMA..6660401C,2024JCAP...09..027Z,Zhang:2024lsf,2024arXiv240907248H}. Accretion disks, as the primary sources of visible light around black holes, significantly influence the resulting images. For the current EHT targets, M87* and Sgr A*, the surrounding hot plasma is considered to be part of a radiatively inefficient accretion flow (RIAF) \citep{2003ApJ...592.1042I,2003PASJ...55L..69N,2014ARA&A..52..529Y,2019ApJ...875L...5E, 2022ApJ...930L..16E,2022MNRAS.511.3795N}. Such flows are typically hot, geometrically thick, and optically thin at an observing frequency of $230~\mathrm{GHz}$, making it possible to observe features close to the event horizon, such as the photon ring, black hole shadow and inner shadow \citep{1973blho.conf..215B, 2000ApJ...528L..13F, 2019PhRvD.100b4018G,2020SciA....6.1310J, 2021ApJ...918....6C, 2021PhRvD.104l4010P}. At an observational frequency of $230~\mathrm{GHz}$, the emission we primarily detected comes from synchrotron radiation produced by relativistic electrons.
However, in the strong gravitational field near black holes, it is not possible to directly infer information about the plasma density, temperature, velocity, or magnetization of accretion disks from the observed images alone. This is typically achieved through general relativistic ray tracing (GRRT) \citep{1979A&A....75..228L,2011CQGra..28v5011V,2016ApJ...820..105P,2021ApJ...912...39G,2022ApJS..262...28W,2024JCAP...11..054H}. By fitting observational data from the EHT to simulated images produced by GRRT, one can effectively extract information about the black hole and the surrounding accretion flow \citep{2020ApJ...897..139B,2021ApJ...910L..13E,2024ApJ...964L..26E,2024MNRAS.535.3181Y}. GRRT involves calculating simulated images of the black hole as seen by the observer, starting with backward ray tracing from the observer to the black hole, followed by radiation transfer calculations through the radiative regions of the accretion disk. However, due to the computationally intensive nature of this process, substantial computational resources are often required. On one hand, to cover a large parameter space, simulations often require the generation of millions of images, demanding significant computational power. On the other hand, Bayesian parameter estimation of the model requires rapid generation of images corresponding to different parameter sets, placing high demands on computational speed \citep{2024MNRAS.535.3181Y, 2024IAUGA..32P1132C}. Most implementations have incorporated GPU acceleration or adaptive ray-tracing methods, yielding significant improvements in computational efficiency \citep{2018ApJ...867...59C,2021ApJ...912...39G,2023ApJS..265...22M}. Recently, advancements in deep neural networks have enabled the generation of black hole images through deep learning, providing a promising new approach for efficiently obtaining such images.

Deep learning models possess powerful feature learning and pattern recognition capabilities, enabling them to extract valuable information from limited data \citep{Krizhevsky2012, Hochreiter1997, 2014Very}. Recent applications of deep learning have shown notable effectiveness in black hole parameter identification \citep{2020A&A...636A..94V,2023MNRAS.520.4867Q} and image reconstruction tasks \citep{2018ApJ...864....7M,2023ApJ...947L...7M}. Several highly effective models have been developed for image generation tasks, offering robust methods for synthesizing realistic, high-quality images.
Generative Adversarial Networks (GANs), which are based on game theory \citep{Generative2014}, use adversarial training to enable two neural networks, namely the generator and the discriminator, to compete with each other, thereby enhancing the generator's ability to produce high-quality samples. While GANs can generate highly realistic images through this adversarial process, the training can be unstable and is prone to mode collapse.
Variational Autoencoders (VAEs) are another class of generative models that learn the probability distribution of data and can generate new samples similar to the training data \citep{kingma2022}. Compared to GANs, VAEs training is more stable; however, VAEs typically produce less realistic images, especially in terms of fine details.
The generation of black hole images had been successfully achieved using both GANs \citep{2024MNRAS.52710965M} and VAEs \citep{2024ApJ...967..140S}.
Compared with GANs and VAEs, diffusion models have a wider range of practical applications \citep{Jonathan2020}. Due to the direct optimization of the logarithmic likelihood of the generation process during training, the diffusion model avoids the common mode collapse problem in GANs. The diffusion model gradually recovers the original data from Gaussian noise through a multi-step denoising process, and the generated samples have very high quality, especially in image generation tasks. The generated images have rich details and high resolution, avoiding the blurring problem of VAEs generated images.

Therefore, we propose a novel black hole image generation method, Branch-Corrected Denoising Diffusion Model (BCDDM), designed to rapidly produce high-quality black hole data.
Diffusion models, as advanced image generation models, have demonstrated their capability to generate high-fidelity images across various domains\citep{10441276, 10570220, zheng2025diffusionbased, tu2024motioneditor, ran2024towards}. However, to the best of our knowledge, no prior research has applied them to black hole image generation. This study aims to explore whether BCDDM can learn the mapping between black hole parameters and images during the generation process. In conventional machine learning tasks involving images, data augmentation techniques such as random rotation and cropping are commonly employed. However, for black hole images, the size and orientation often carry specific physical significance, rendering such operations unsuitable as they may distort the interpretability of the data. BCDDM addresses this limitation by providing an alternative augmentation strategy—generating physically plausible synthetic images conditioned on parameter inputs, thereby expanding the training dataset while preserving physical consistency. To validate the utility of BCDDM-generated images, we train a ResNet50 regressor on both the original dataset and the augmented dataset produced by our model, aiming to demonstrate improved predictive performance for black hole parameters.

The main contributions of this article are as follows:
Firstly, this paper presents an innovative diffusion architecture model, BCDDM, which addresses the limitations of traditional methods in accurately mapping images to physical parameters during black hole image generation. By incorporating parameter correction branches and a mixed loss function, BCDDM significantly enhances the precision and diversity of the generated images.
Additionally, BCDDM is the first to apply diffusion algorithms to black hole image generation. The experimental results demonstrate that the model is capable of producing high-quality black hole images with a strong correlation to physical parameters, thereby offering novel tools and methodologies for astronomical observations and theoretical studies.
Finally, the integration of a ResNet50-based black hole parameter regressor confirms the high fidelity of images produced by BCDDM and underscores the model's substantial contribution to improving the performance of data-driven regressors in parameter prediction.

\section{Methodology}

\subsection{Black hole image dateset}

We constructed a dataset using the radiatively inefficient accretion flow (RIAF) model \citep{1997ApJ...476...49N, 2014ARA&A..52..529Y}, focusing exclusively on thermal synchrotron emission from the accretion flow. In such an accretion flow, the magnitude and distribution of the electron temperature and electron density directly affect the distribution and intensity of the radiation in the black hole image. The spatial distributions of the electron temperature \( \mathcal{T}_e \) and thermal electron density \( \mathcal{N}_e \) are described by
\begin{equation}
    \mathcal{N}_e = n_{e} \, r^{-\alpha} e^{-\beta}, \quad \mathcal{T}_e = T_{e} \, r^{-\gamma}, \quad \beta = \frac{z^2}{2\sigma^2},
\end{equation}
where \( z \equiv r \cos{\theta} \), and the radial dependencies \( \alpha = 1.1 \) and \( \gamma = 0.84 \) are adopted from vertically averaged density and temperature profiles \citep{2003ApJ...598..301Y,2016ApJ...831....4P,2018ApJ...863..148P}. The electron number density \( n_e \) and temperature \( T_e \) define the normalization of the electron distribution. The parameter \( \sigma \) controls the vertical thickness of the accretion flow, parameterized as \( h_{\mathrm{disk}} \equiv \sigma / x \), where \( x \equiv r \sin{\theta} \). The geometry of the flow is thus determined by \( h_{\mathrm{disk}} \).
The magnetic field strength \( B \) is assumed to be in approximate equipartition with the ions
\begin{equation}
    \frac{B^2}{8\pi} = \epsilon n_e \frac{m_p c^2 r_g}{6 r},
\end{equation}
where \( r_g = G M_{\mathrm{BH}} / c^2 \), \( m_p \) is the proton mass, and \( \epsilon = 0.1 \) \citep{2011ApJ...735..110B,2011ApJ...738...38B,2016ApJ...820..137B,2016ApJ...831....4P,2018ApJ...863..148P}. The Keplerian factor \( k \) describes the degree to which the flow deviates from Keplerian motion (\( k = 1 \)) or approaches free-fall (\( k = 0 \)) \citep{2016ApJ...831....4P}. A key limitation of the RIAF model is the lack of jet emission. For M87*, EHT observations at $230~\mathrm{GHz}$ show significant synchrotron contributions from both the disk and the jet on horizon scales \citep{2019ApJ...875L...1E, 2019ApJ...875L...5E}. This omission in the model could be addressed in the future by employing more comprehensive models, such as GRMHD simulations.
We generate the dataset using the GRRT code \textit{ipole} \citep{2018MNRAS.475...43M}, which calculates the radiation intensity distribution for simulated observational images. These images are produced at a frequency of $230~\mathrm{GHz}$, with a camera field of view of \( 160 \times 160 \ \mu \mathrm{as} \) and resolution of \( 256 \times 256 \). To align with the M87* case, we fix the observer inclination angle at \( 163^\circ \) and adjust the total flux density to \( 0.5 \ \mathrm{Jy} \) \citep{2019ApJ...875L...5E} by adjusting \( n_e \). Seven physical parameters are varied in the simulation, with their respective value ranges provided in Table \ref{tab:parameter}. The rotation direction of the accretion disk is randomly selected between \( -1 \) (clockwise) and \( 1 \) (counterclockwise), while the other six parameters are uniformly sampled within their respective ranges.
Ultimately, we constructed a dataset of $2157$ images for training the BCDDM. These datasets are openly accessible to support further research and validation\footnote{The datasets used in this study are publicly available at \url{https://doi.org/10.5281/zenodo.15354648}.}.

\begin{table}[ht]
     \caption{\centering{Parameter ranges used in the RIAF black hole images generation.}}
    \centering
    \begin{tabular}{|c|c|c|}
        \hline
        \textbf{Parameter} & \textbf{Symbol} & \textbf{Range} \\
        \hline
        Black hole spin & \(a\) & \([-1, 1]\) \\
        Black hole mass & \(M_{\mathrm{BH}}\) & \([5 \times 10^9 M_{\odot}, 8 \times 10^{9} M_{\odot}]\) \\
        Electron temperature & \(T_e\) & \([10^{10} \, \mathrm{K}, 10^{12} \, \mathrm{K}]\) \\
        Accretion disk thickness & \(h_{\mathrm{disk}}\) & \([0.1, 0.8]\) \\
        Keplerian factor & \(k\) & \([0, 1]\) \\
        Position angle & \(PA\) & \([0^\circ, 360^\circ]\) \\
        Fluid direction & \(F_{\mathrm{dir}}\) & \(-1 \) ,\( 1\) \\
        \hline
    \end{tabular}
    \label{tab:parameter}
\end{table}

\subsection{Data preprocessing}
The parameters used for generating black hole images exhibit significant disparities in their orders of magnitude. As shown in Table \ref{tab:parameter}, certain parameters like black hole spin and mass differ by up to $9$ orders of magnitude.
In deep learning frameworks, data covering very different orders of magnitudes in their parameter space can lead to numerical computation problems such as gradient explosion or vanishing, which can seriously affect the training effectiveness and convergence performance of the model. To alleviate these numerical instabilities, we introduce Z-score normalization technique to preprocess black hole parameters \citep{Fei2021}. Z-score quantifies the standard deviation of a single data point relative to the mean of the entire dataset. Through Z-score, the raw data is transformed into a standard normal distribution, with its mean adjusted to $0$ and standard deviation to $1$, ensuring consistency in data scale. Let $r$ denote the parameter category, where $x_r$ represents the initial black hole parameter value, $\mu_r$ is the mean of the initial black hole parameters, and $\sigma_r$ is their standard deviation. The normalized black hole parameter value $z_r$ is given by the Z-score formula
\begin{equation}
     z_{r} = \frac{x_{r} - \mu_{r}}{\sigma_{r}}, \quad \text{for} \quad r \in \{1, 2, \ldots, 7\}.
\end{equation}
where \( T_e \) was log-transformed prior to normalization. Z-score normalization is a reversible linear transformation, allowing the original black hole parameters to be recovered from their normalized counterparts.
To ensure training stability, we also apply Maximum Normalization to the black hole images. For each image, given the flux density \( S(i,j) \) of each pixel, the normalized image \( S_{\text{norm}} \) is computed as
$S_{\text{norm}}(i, j) = S(i, j)/S_{\text{max}}$, where $ S_{\text{max}} = \max(S)$.
This scales the pixel values to the range [0, 1]. Since all black hole images are adjusted to a total flux density of \( 0.5 \, \mathrm{Jy} \), the normalized images generated by the neural network can be rescaled back to their original physical dimensions by restoring the total flux density to \( 0.5 \, \mathrm{Jy} \).

\subsection{Diffusion model framework}

\begin{figure}[ht!]
     \centering
     \includegraphics[width=1\linewidth]{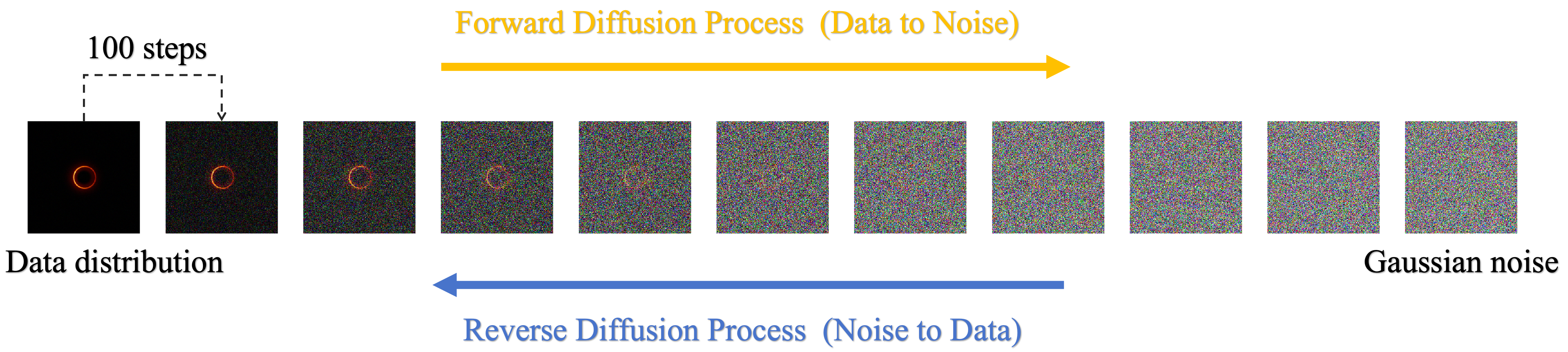}
     \caption{\label{fig-3} The diffusion process of black hole images. We output the image every $100$ steps for a total of $1000$ times to observe the changes in the diffusion process of the image. }
\end{figure}

The diffusion model initiates with a forward diffusion process applied to the black hole image, wherein Gaussian noise is incrementally introduced according to a predefined noise schedule. This process systematically corrupts the input image over \(T\) steps until it converges to a pure Gaussian noise distribution. The noise schedule, governed by the diffusion coefficients \(\beta_t\), determines the magnitude of noise added at each step \(t\). Specifically, the transition from \(x_{t-1}\) to \(x_t\) is formulated as
\begin{equation}
     q(x_t | x_{t-1}) = \mathcal{N}(x_t; \sqrt{1 - \beta_t} x_{t-1}, \beta_t I),
\end{equation}
which can equivalently be expressed as
\begin{equation}
     x_t = \sqrt{1 - \beta_t} x_{t-1} + \sqrt{\beta_t} \epsilon_t,
\end{equation}
where \(\epsilon_t \sim \mathcal{N}(0,1)\). The joint distribution of the entire forward process is given by
\begin{equation}
     q(x_{1:T} | x_0) = \prod_{t=1}^T q(x_t | x_{t-1}).
\end{equation}
As illustrated in Figure \ref{fig-3}, the diffusion process is visualized for \(T = 1000\) steps, with \(\beta_t\) initialized at \(10^{-4}\) and linearly increasing to \(0.02\). The linear noise schedule ensures a smooth degradation of the black hole image into noise, balancing detail preservation in early stages with complete corruption in later stages.

During the training process, the model learns to reverse this process by predicting the noise introduced at each diffusion step. The denoising process involves estimating the conditional distribution \(q(x_{t-1} | x_t, x_0)\), which is Gaussian with mean \(\mu_t\) and variance \(\beta_t\)
\begin{equation}
     q(x_{t-1} | x_t, x_0) = \mathcal{N}(x_{t-1}; \mu_t, \beta_t I).
\end{equation}
By defining \(\alpha_t = 1 - \beta_t\) and \(\bar{\alpha}_t = \prod_{s=1}^t \alpha_s\), the mean and variance can be derived as
\begin{equation}
     \mu_t = \frac{\alpha_t}{\sqrt{1 - \alpha_t}} \left ( x_t - \frac{\sqrt{1 - \bar{\alpha}_t}}{\sqrt{1 - \alpha_t}} \epsilon_t \right )
\end{equation}
\begin{equation}
     \beta_t = \frac{1 - \bar{\alpha}_{t-1}}{1 - \bar{\alpha}_t} \beta_t
\end{equation}
This formulation enables the trained noise-prediction network to iteratively denoise random Gaussian noise, ultimately reconstructing a coherent black hole image from pure noise. The controlled and gradual nature of the diffusion process is critical for ensuring the model's ability to accurately reverse the noise addition, thereby facilitating high-quality sample generation.

\subsection{Model architecture}

\begin{figure}[ht!]
     \centering
     \includegraphics[width=1\linewidth]{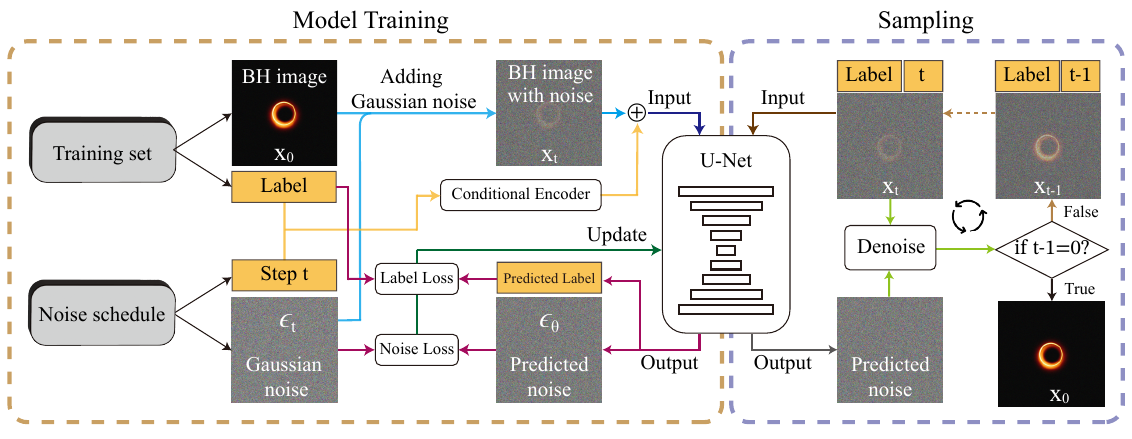}
     \caption{\label{fig-1} BCDDM architecture for black hole image generation. The diffusion process and denoising process utilize the same U-Net model, with detailed architecture illustrated in Figure \ref{fig-2}.}
\end{figure}

\begin{figure}[ht!]
     \centering
     \includegraphics[width=1\linewidth]{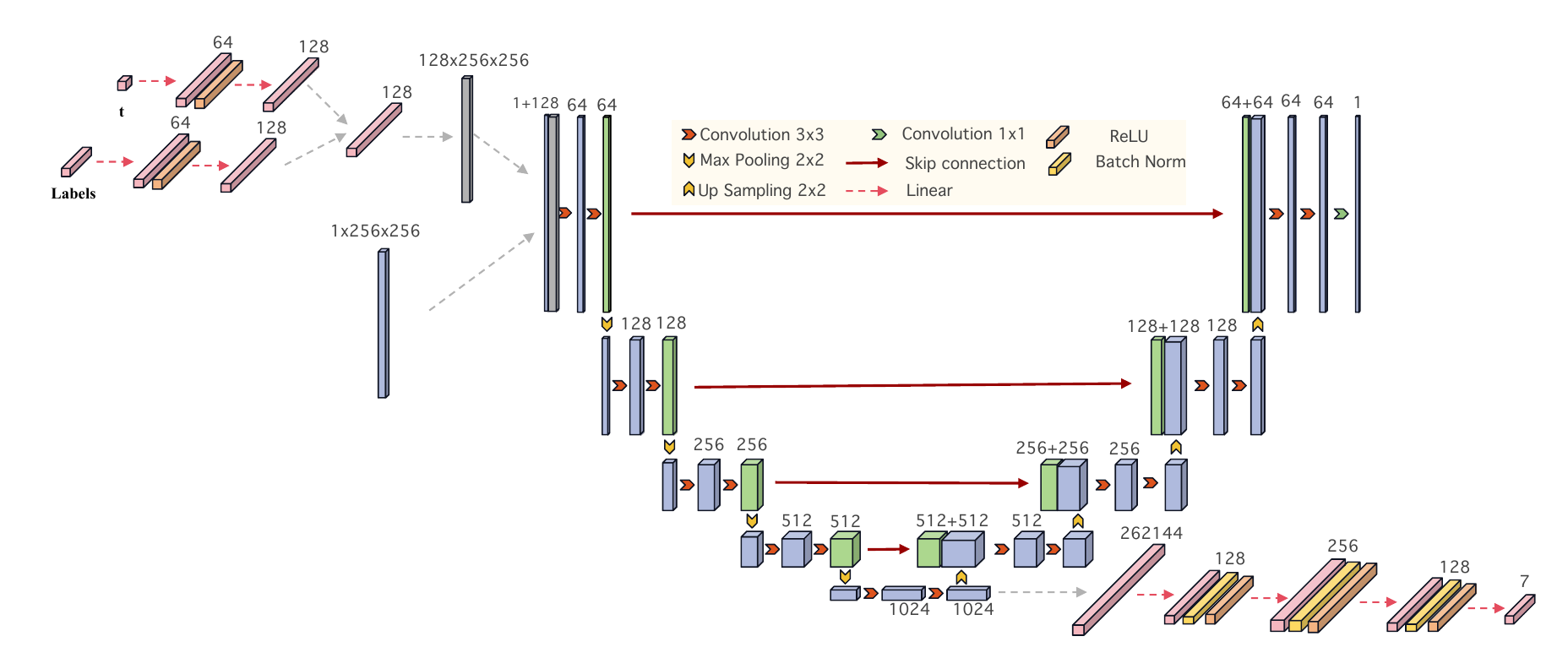}
     \caption{\label{fig-2} Encoding and Decoding architecture embedded in BCDDM. The trainable components consist solely of the input encoded label vector and black hole image, along with the output predicted parameters and noise.}
\end{figure}

The logical architecture of BCDDM for its training and sampling to generate images is shown in Figure \ref{fig-1}. Our training set contains the black hole images we generated, and the seven physical parameters for generating the black hole images serve as the labels for each image. During training, the forward diffusion process is first applied to the original black hole image \(x_0\), progressively introducing noise according to the noise schedule. After \(t\) steps of noise addition, the resulting image \(x_t\) is obtained. To establish a correspondence between the black hole image and its physical parameters, we encode both the step \(t\) and the image's label, embedding this information into the noisy image \(x_t\). The core of our proposed model lies in a Branch-Corrected U-Net architecture designed to perform physically conditioned denoising, enabling the restoration of high-fidelity black hole images that align with the physical parameters. As shown in Figure \ref{fig-2}, we enhance the standard U-Net architecture by introducing a parameter corrector branch at the lowest point of the downsampling path. This branch encodes the extracted features to learn the mapping between the image's high-dimensional latent space distribution and the physical parameters.
The Branch-Corrected U-Net simultaneously predicts both the noise component \(\epsilon_\theta\) and the parameter label. The model parameters are updated by minimizing the residuals between the predicted and actual noise, as well as between the predicted and ground truth labels, ensuring both denoising accuracy and physical fidelity.

To optimize the model, we design a composite loss function combining noise prediction loss \(L_{\text{noise}}\) and parameter consistency loss \(L_{\text{label}}\) \citep{Wan2000}. The noise loss, defined as
\begin{equation}
L_{\text{noise}} = \mathbf{E}_{t, x_0, \epsilon} \left[ \left( \epsilon - \epsilon_{\theta}(x_t, t, l) \right)^2 \right],
\end{equation}
ensures high-quality image generation by penalizing deviations in noise estimation. Meanwhile, the label loss,
\begin{equation}
L_{\text{label}} = \mathbf{E}_{x_0, l} \left[ \left( f(k) - l \right)^2 \right],
\end{equation}
where \(k\) denotes the feature vector extracted during downsampling and \(f(k)\) its encoded representation, enforces alignment between generated images and their target physical parameters. The total loss is a weighted combination of these terms,
\begin{equation}\label{eq:mixd_loss}
L = \lambda_1 L_{\text{noise}} + \lambda_2 L_{\text{label}},
\end{equation}
where \(\lambda_1\) and \(\lambda_2\) balance the contributions of denoising and physical consistency during optimization. This dual-objective optimization ensures accurate denoising while preserving physical consistency.

Once the model training is completed, the reconstruction of a black hole image from input parameters can be achieved through a sampling process, as illustrated in Figure \ref{fig-1}. The process begins by initializing the step $t = T$ and providing an initial Gaussian noise $x_T$ along with the physical parameters of the black hole image as the label. These inputs are fed into the model, which predicts the noise component and subsequently denoises $x_T$ to produce the image $x_{T-1}$ at step $t = T-1$. This iterative procedure is repeated, progressively reducing the noise at each step, until the final step $t = 0$ is reached. At this stage, the sampling process concludes, yielding the fully denoised black hole image $x_0$ conditioned on the physical parameter labels.

\section{Results}
\subsection{Generate Black Hole Image Results}

The BCDDM model was trained on the RIAF black hole image dataset, which was partitioned into $1941$ images for training and $216$ images for validation. The noise schedule was configured with a total diffusion step $T = 1000$, where the diffusion coefficient \(\beta_t\) was initialized at \(10^{-4}\) and linearly increased to \(0.02\). For optimization, we fixed the weighting coefficients in Eq. \ref{eq:mixd_loss} as $\lambda_1 = 0.95$ and $\lambda_2 = 0.05$ throughout the training process. This configuration ensured the validation loss values of the noise and the label with a small difference, and they will ultimately be $5.27\times 10^{-4}$ and $4.13 \times 10^{-4}$, respectively. The learning rate was initially set to $3 \times 10^{-3}$ and adjusted using a loss-driven decay strategy to improve model convergence. Specifically, after $500$ epochs, if no improvement in validation loss was observed for $5$ consecutive epochs, the learning rate was reduced by a factor of $0.97$. This annealing process resulted in a final learning rate of approximately $1.07 \times 10^{-5}$, ensuring stable training dynamics while avoiding premature convergence to suboptimal solutions.
\begin{figure}
     \centering
     \includegraphics[width=0.6\linewidth]{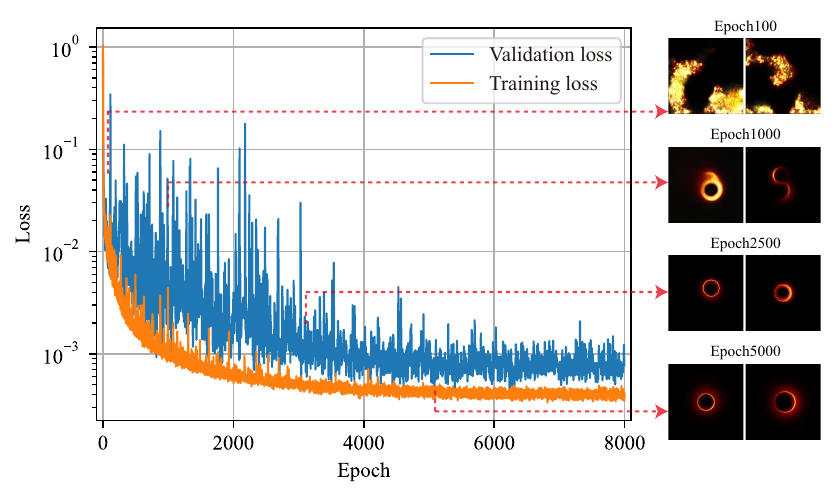}
     \caption{\label{fig4} Training loss and validation loss during the training process. The right panel displays two independent black hole images generated by BCDDM at epochs $100$, $1000$, $2500$, and $5000$, illustrating the performance progression of the generative model.}
\end{figure}

The model was trained for $8000$ epochs, with the loss trajectory depicted in Figure \ref{fig4}. The initial phase exhibited a rapid decline in loss, followed by a slower convergence in later stages. Notably, the validation loss plateaued after $5000$ epochs, while the training loss continued to decrease slowly, indicating the onset of overfitting. To illustrate the model's performance progression, two independent black hole images were sampled at epochs $100$, $1000$, $2500$, and $5000$. The results demonstrate a clear correlation between decreasing loss and improving image clarity, with the generated images becoming progressively sharper as training advanced.

\begin{figure}
     \centering
     \includegraphics[width=0.95\linewidth]{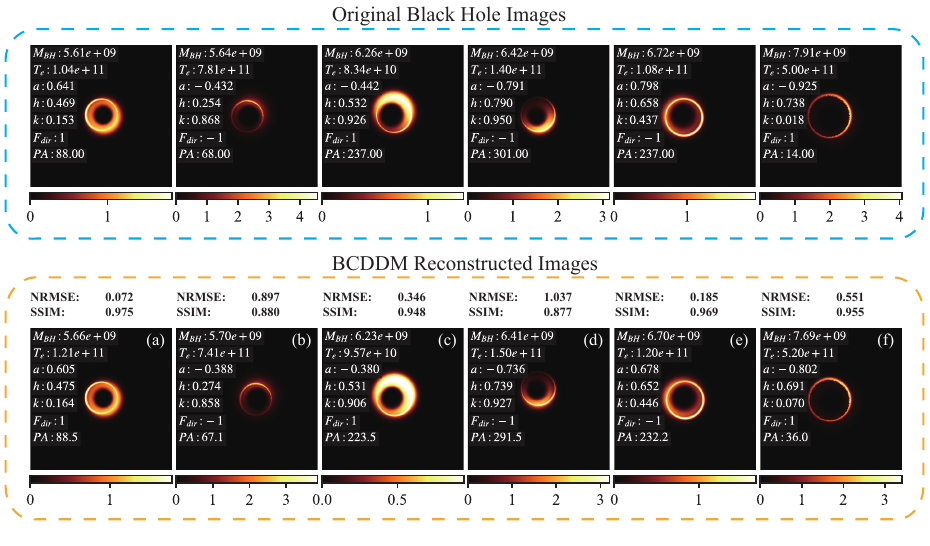}
     \caption{\label{fig5} Comparison of original and reconstructed black hole images in the additional test set. The figure presents six pairs of GRRT-simulated original black hole images (top row) alongside their BCDDM-reconstructed counterparts (bottom row). The parameters displayed above the original images represent the ground truth values used as initial inputs, while those below the reconstructed images show the output from the parameter corrector branch. Each reconstructed image is annotated with its NRMSE (Normalized Root Mean Square Error) and SSIM (Structural Similarity index) values compared to the original. All images are rendered in brightness temperature units ($10^{10}K$), calculated as $T = S\lambda^2/(2k_B\Omega)$, where $S$ denotes flux density, $\lambda$ the observation wavelength, $k_B$ the Boltzmann constant, and $\Omega$ the solid angle of the resolution element.}
\end{figure}

Using the model saved at the 5000th training epoch, we generate images from physical parameters and Gaussian noise. We select $6$ images from an additional test dataset and reconstruct the corresponding black hole images using their associated parameter sets, as shown in Figure \ref{fig5}. In the original GRRT image, the initial parameters used for generating the image are overlaid for reference. Meanwhile, the BCDDM-reconstructed image displays the predicted output from the parameter correction branch. The results demonstrate that BCDDM produces images with consistent morphological and brightness distribution features compared to those generated by GRRT when using the same physical parameters. Key features such as the direct image of the accretion disk, photon ring, and inner shadow are accurately reconstructed. To quantify the similarity between the reconstructed and original images, we compute the Normalized Root Mean Squared Error (NRMSE) \citep{2014GeopP..62.1009S} and the Structural Similarity Index (SSIM) \citep{4775883, 1284395}, displayed above each BCDDM-reconstructed image. NRMSE measures pixel-wise similarity, while SSIM emphasizes structural consistency, where an NRMSE of $0$ and an SSIM of $1$ indicate perfect agreement. Our analysis reveals high SSIM values across all reconstructed images. However, panels (b) and (d) in Figure \ref{fig5} exhibit relatively higher NRMSE values. This discrepancy arises because, despite the consistency in image features, the spatial alignment between BCDDM-generated and GRRT images occasionally varies, likely due to the influence of initial noise distribution during sampling from pure Gaussian noise. Additionally, differences in the maximum values of the color bars indicate slight instability in the model’s brightness prediction, contributing to elevated NRMSE values, as seen in panels (c) and (f).
The physical parameters predicted by the parameter correction branch closely match the original values, with only minor discrepancies observed in the spin parameter $a$ and electron temperature $T_e$. This confirms that the parameter correction branch effectively learns the mapping between images and physical parameters, demonstrating BCDDM’s sensitivity to parameter variations.

\begin{figure}
     \centering
     \includegraphics[width=0.95\linewidth]{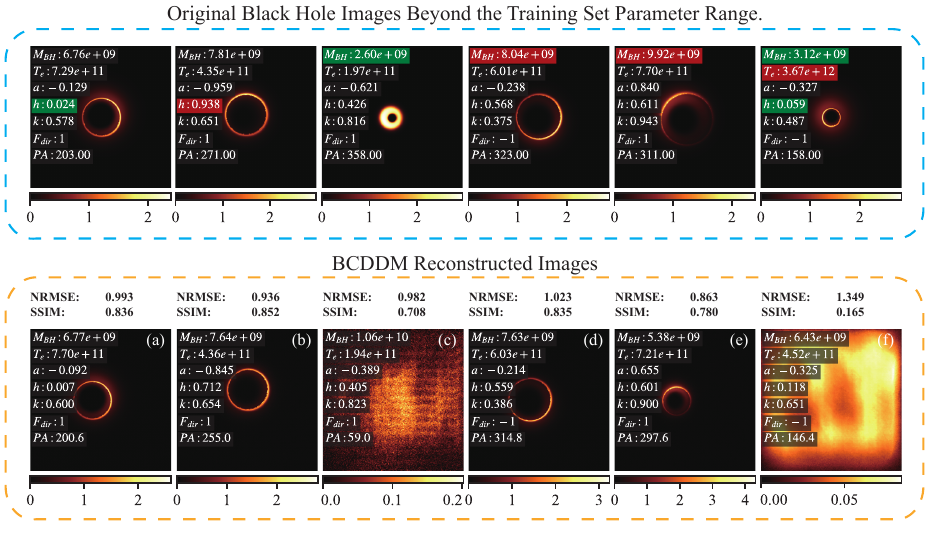}
     \caption{\label{fig-7} Comparison of the original and reconstructed black hole images, similar to Figure~\ref{fig5}, but evaluated beyond the training dataset's parameter range. Parameters marked in green indicate values below the original range, while those in red denote values above the range.}
\end{figure}

While we selected an extensive parameter range for $a$, $k$, $F_{\text{dir}}$, and PA during training, the parameters $M_{\text{BH}}$, $T_e$, and $h$ could potentially span even wider ranges. The model's ability to reconstruct images beyond the original dataset's parameter range provides insight into its generalization capabilities. Figure \ref{fig-7} presents six experiments where specific parameters were either increased or decreased beyond the original bounds. The results indicate that the model only achieves partial generalization for the accretion disk thickness h, successfully recovering the image structure as shown in panels (a) and (b). However, the model shows limited generalization for $M_{\text{BH}}$ and $T_e$, while minor deviations from the training range yield consistent reconstructions (panel d), larger deviations result in either failed reconstructions (panels c and f) or artifacts in recovered features (panel e). All reconstructed images exhibit high NRMSE values, suggesting that the current model primarily specializes in the trained parameter space. Nevertheless, this limitation could be mitigated by expanding the parameter range of the training dataset.

\subsection{Parameter Regression Evaluator}
\begin{table}
     \caption{\label{tab1}\centering{Dataset size and split for RLDs, FKDs, and MXDs.}}
     \centering
     \resizebox{0.3\textwidth}{!}{
     \begin{tabular}{cccc}
     \hline
     Dataset & Train & Val & Test\\
     \hline
     RLDs & 1725 & 216 & 216   \\
     FKDs & 1725 & 216 & 216   \\
     MXDs & 3450 & 216 & 216   \\
     \hline
     \end{tabular}
     }
\end{table}

To evaluate the effectiveness of BCDDM in expanding the original RIAF black hole image dataset, we construct a deep learning-based black hole parameter regression model. This model is trained on both the original dataset and the dataset augmented by BCDDM. By comparing the model's performance on these two datasets, we aim to determine whether the augmented dataset can enhance the prediction accuracy of the black hole parameter regression model.

The RIAF dataset serves as the foundation for constructing the real dataset (RLDs), which includes training, validation, and test sets. We then generate an equivalent number of black hole images using BCDDM through uniform random sampling within the parameter ranges specified in Table \ref{tab:parameter}. Using a single RTX3090 graphics card, we can generate images at a speed of $5.25$ seconds per image. The resulting datasets comprise three distinct configurations: the fake dataset (FKDs) containing exclusively BCDDM-generated images, the original RLDs, and a mixed dataset (MXDs) combining both RLDs and FKDs. To maintain evaluation consistency, both FKDs and MXDs utilize the same validation and test sets as RLDs. The respective sizes of these datasets are detailed in Table \ref{tab1}.

\begin{figure}[ht!]
     \centering
     \includegraphics[width=0.85\linewidth]{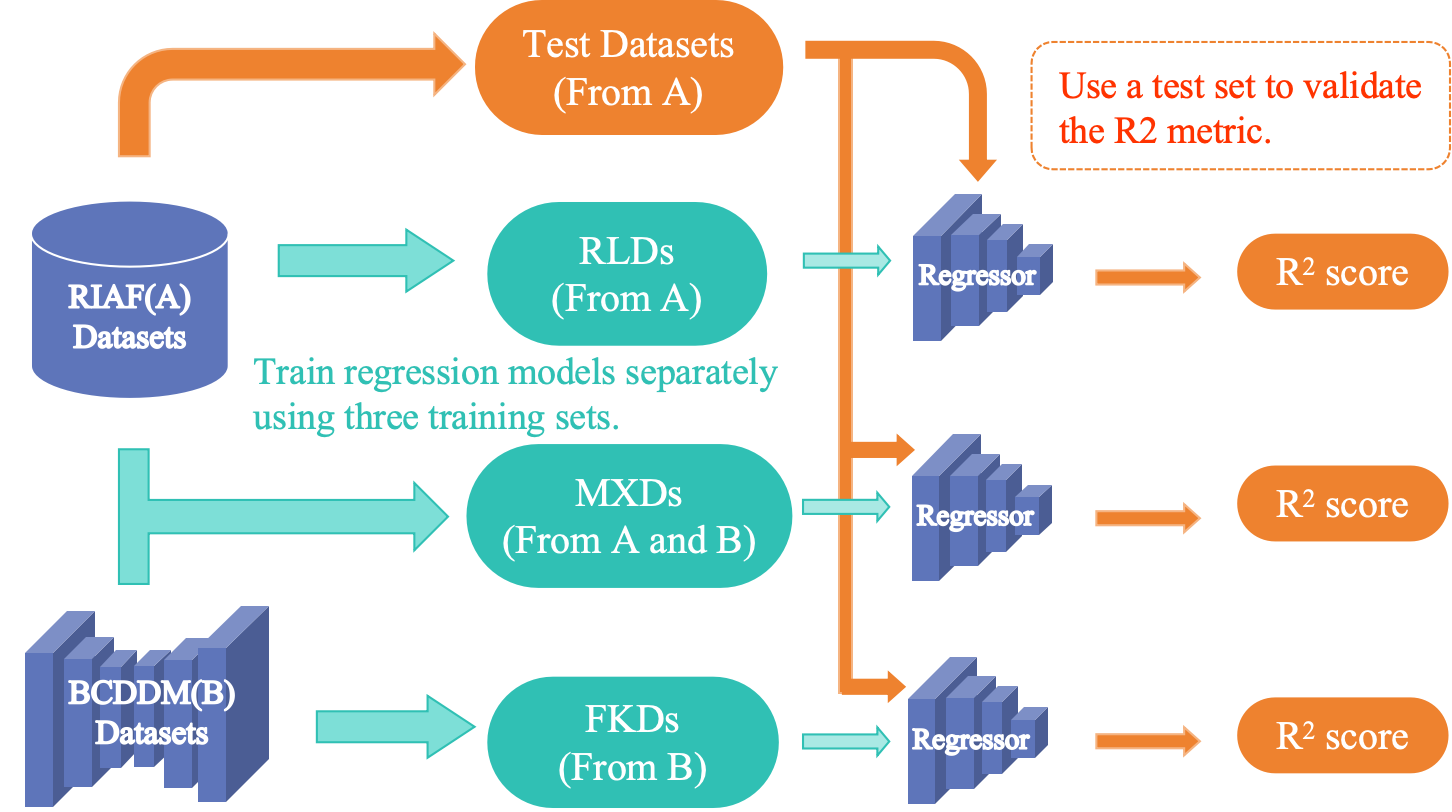}
     \caption{\label{fig-8} Workflow of Constructing Datasets from Multiple Sources and Evaluating Regression Model Performance. This figure outlines the process of constructing datasets from diverse sources and evaluating the performance of a regression model. It highlights two primary data sources: RIAF Datasets and BCDDM Datasets. The figure underscores the critical step of using a test set to validate the $R^2$ metric, culminating in the output of the $R^2$ score to assess the model's performance.}
\end{figure}

As illustrated in Figure \ref{fig-8}, we evaluate the performance of three ResNet50 based regressors \citep{He2016} using different training sets. We employ $R^2$ as a quantitative measure to assess improvements in parameter estimation accuracy. The $R^2$ metric is widely used in regression analysis in machine learning. A value closer to $1$ indicates a better fit of the model to the data. Let \( y_i \) denote the true parameter value of the \( i \)-th sample of the test set, \( \hat{y}_i \) represent the model's predicted value for the \( i \)-th sample, \( \bar{y} \) be the mean parameter value across the test set, and \( n \) denote the total number of test samples. The \( R^2 \) score is calculated as
\begin{equation}
    R^2 = 1 - \frac{\sum_{i=1}^{n} (y_i - \hat{y}_i)^2}{\sum_{i=1}^{n} (y_i - \bar{y})^2}.
\end{equation}

\begin{table}
     \caption{\label{tab2}\centering{$R^2$ values for different parameters in three datasets.}}
     \centering
     \resizebox{\textwidth}{!}{
     \begin{tabular}{ccccccc}
     \hline
     Trainset & $a$ & $T_{e}$ & $h_{disk}$ & $M_{BH}$ & $k$  & $PA$\\
     \hline
     RLDs & 0.9206 & 0.9671 & 0.9841 & 0.9982 & 0.9538  & 0.9012 \\
     FKDs & 0.7992($\downarrow$0.1214) & 0.9793($\uparrow$0.0122) & 0.9683($\downarrow$0.0158) & 0.9913($\downarrow$0.0069) & 0.9157($\downarrow$0.0381) & 0.7789($\downarrow$0.1223) \\
     MXDs & 0.9645($\uparrow$0.0439) & 0.9960($\uparrow$0.0289) & 0.9893($\uparrow$0.0052) & 0.9982($-$0.0000) & 0.9762 ($\uparrow$0.0224)& 0.9602($\uparrow$0.0590) \\
     \hline
     \end{tabular}
     }
\end{table}

The Table \ref{tab2} shows the $R^2$ scores of three datasets experiments. We observe that although the FKDs composed of synthesized images scored lower than RLDs, the model trained on reconstructed images is acceptable for predicting RIAF black hole images. The more important point is that the  $R^2$  scores of MXDs are substantially higher for each parameter compared to those of RLDs. This suggests that incorporating synthetic images exposes the model to a broader spectrum of feature combinations during training, thereby enhancing both the diversity and volume of training data. Consequently, this approach effectively mitigates the risk of model overfitting. Meanwhile, the increase in $R^2$ value also proves in reverse that BCDDM can learn the mapping relationship between black hole images and black hole parameters. The synthesized images generated by BCDDM not only visually resemble real images, but also have a high degree of consistency in statistical properties and physical meaning. This result demonstrates the feasibility of BCDDM in synthesizing black hole images. By generating high-quality synthetic images, BCDDM provides a new data augmentation method for black hole research, which helps overcome the problem of data scarcity and promotes the development of astrophysics.

\begin{figure}[ht!]
     \centering
     \includegraphics[width=0.85\linewidth]{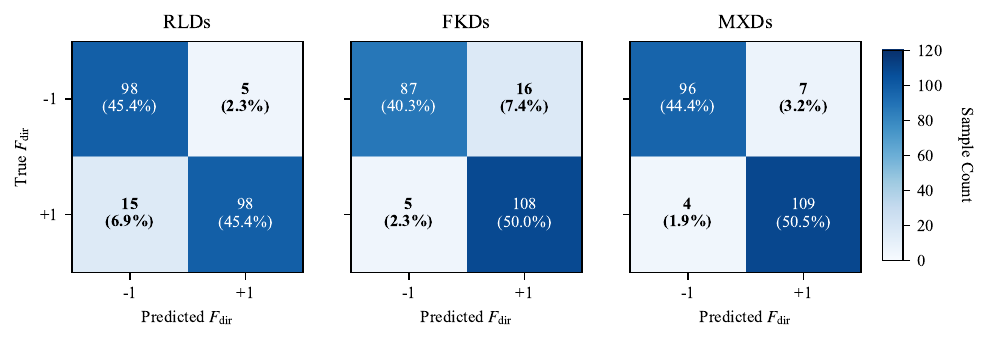}
     \caption{\label{fig9}The binary confusion matrix of $F_{dir}$. The datasets corresponding to the three images are RLDs (left), FKDs (middle), and MXDs (right). The classification accuracies of three training sets for fluid-direction are 92.19\%, 89.58\% and 94.27\%. }
\end{figure}

\begin{figure}
     \centering
     \includegraphics[width=0.85\linewidth]{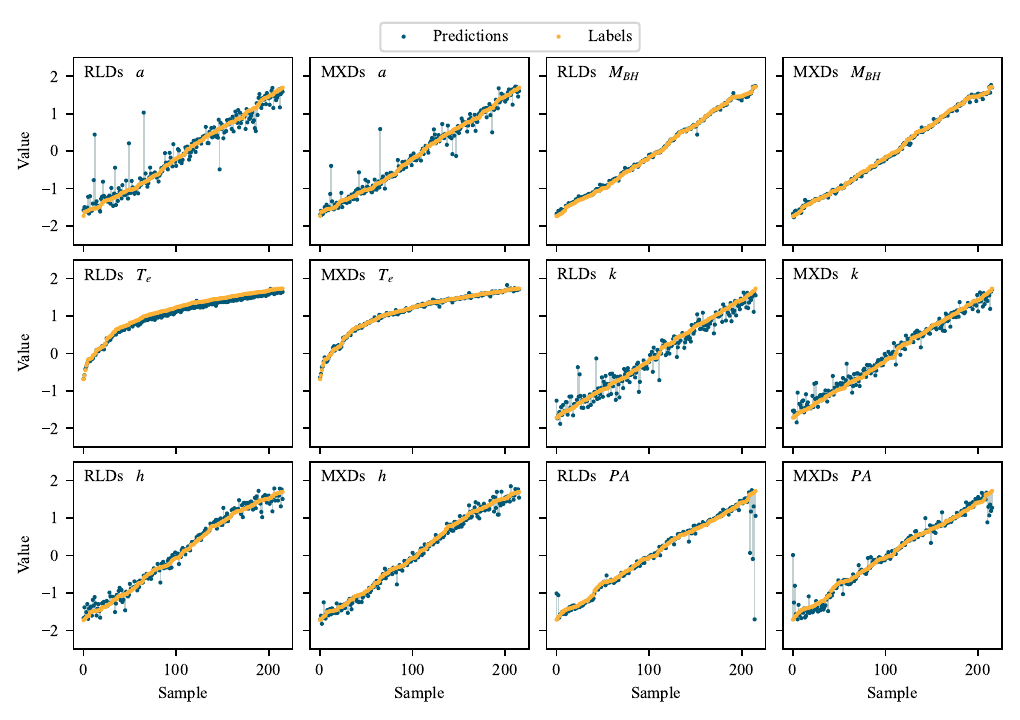}
     \caption{\label{fig10}The fitting results of RLDs and MXDs on the test set samples.The prediction error of the test set after training with RLDs for 5000 epochs (left) and the prediction error of the test set after training with MXDs for 5000 epochs (right). }
\end{figure}

At the same time, as shown in Figure \ref{fig9}, the confusion matrix for binary classification of fluid-dirction also indicates that MXDs has the best classification performance. This indicates that our method can perform data augmentation on small-scale datasets, increasing the size of the training dataset and helping the regressor to more accurately understand the mapping relationship between black hole parameters and images.

In addition, in Figure \ref{fig10}, we show the fitting results of the test set samples for RLDs and MXDs. We sorted the label points by numerical value from small to large and plotted the regressor's predictions and errors for the samples. We can clearly see that compared to RLDs, the predicted results of MXDs are closer to the true values.

\begin{table}
     \caption{\label{tab3}\centering{$R^2$ values of blurred datasets}}
     \centering
     \resizebox{0.8\textwidth}{!}{
     \begin{tabular}{ccccccc}
     \hline
     Trainset & $a$ & $T_{e}$ & $h_{disk}$ & $M_{BH}$ & $k$  & $PA$\\
     \hline
     RLDs(10$\mu$as) & 0.9263 & 0.9756 & 0.9717 & 0.9950 & 0.9751  & 0.8981 \\
     MXDs(10$\mu$as) & 0.9509 $\uparrow$ & 0.9901 $\uparrow$ & 0.9737 $\uparrow$ & 0.9962 $\uparrow$ & 0.9774 $\uparrow$ & 0.8070 $\downarrow$ \\
     RLDs(20$\mu$as) & 0.6893 & 0.9021 & -0.0683 & 0.9585 & 0.9598 & 0.8417 \\
     MXDs(20$\mu$as) & 0.7732 $\uparrow$ & 0.9799 $\uparrow$ & -0.3325 $\downarrow$ & 0.9865 $\uparrow$ & 0.9648 $\uparrow$& 0.8726 $\uparrow$ \\
     \hline
     \end{tabular}
     }
\end{table}

\begin{figure}
     \centering
     \includegraphics[width=0.85\linewidth]{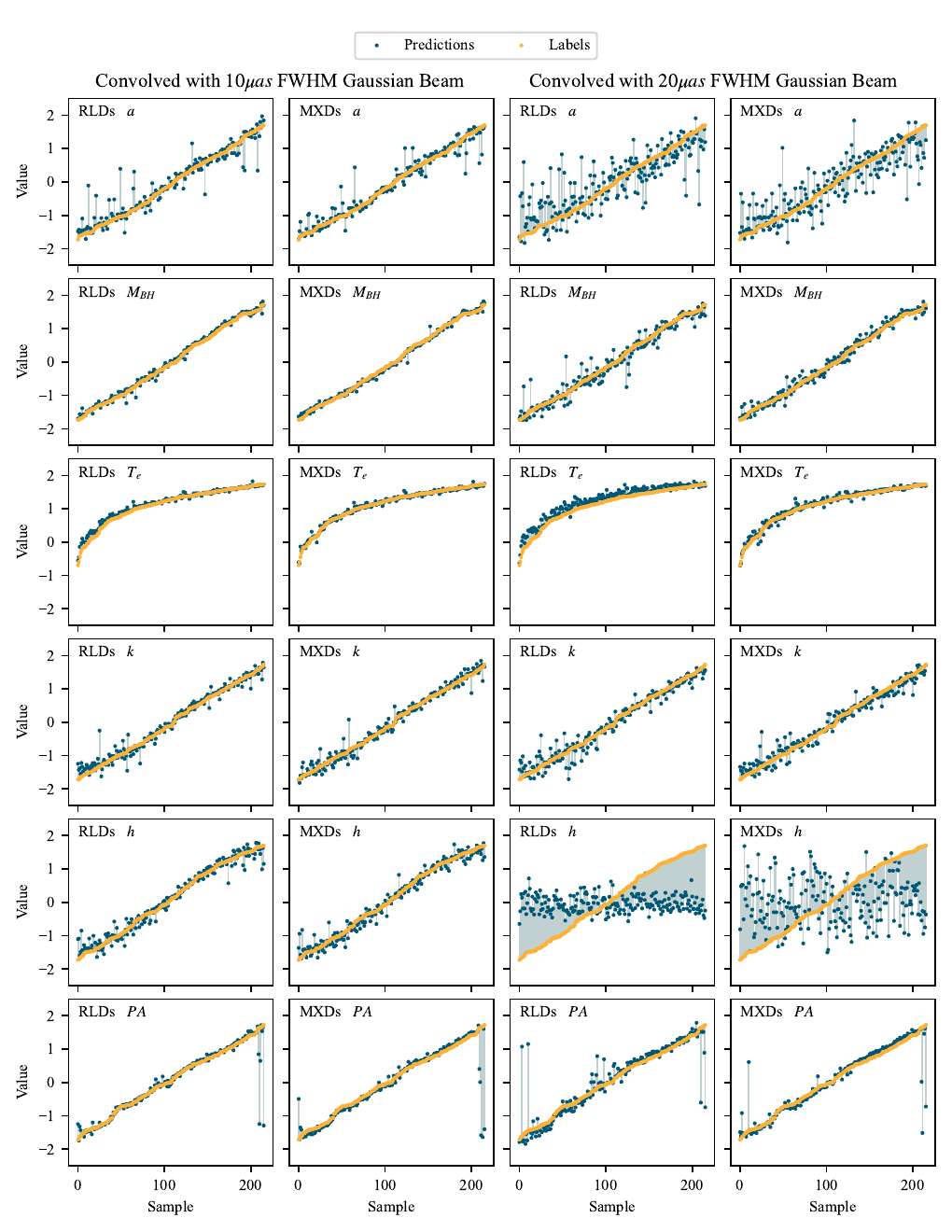}
     \caption{\label{fig11}The fitting results of RLDs and MXDs on the test set samples. The prediction error of the test set after training with RLDs for 5000 epochs (left) and the prediction error of the test set after training with MXDs for 5000 epochs (right). }
\end{figure}

Due to the limited resolution of actual observations, we evaluated the robustness of black hole parameter predictions by blurring the images. As shown in Table \ref{tab3}, we blurred the images using Gaussian kernels with FWHM of $10 \mu as$ and $20 \mu as$, respectively. The experimental results quantitatively reveal three phenomena: when blurring to 10 $\mu$as,  the regressor maintains high accuracy in predicting each parameter, indicating that there is no significant loss of various features of the black hole at this scale. At the present EHT resolution of 20 $\mu$as, the prediction accuracy of the accretion disk height $h_{\mathrm{disk}}$ drops sharply, and the $R^2$ of RLDs decreases from 0.9717 (at 10 $\mu$as) to -0.0683, demonstrating the degradation of accretion disk thickness features. In Figure \ref{fig11}, we present the regression results of blurred datasets. Overall, MXDs show clear advantages over RLDs. We have ample reason to believe that our proposed method can accurately generate clear black hole images while preserving the characteristics of black hole parameters. This achievement is expected to provide strong support and assistance for other data-driven algorithms.

\section{Conclusion}
In this study, we propose an innovative black hole image generation method named Branch Correction Denoising Diffusion Model, aimed at effectively enhancing limited scale black hole image data stes and accelerate the generation of black hole images. By innovatively improving the architecture of the classical diffusion model, we introduce a parameter correction branch and use a mixed loss function to ensure that the model can learn both image distribution and black hole parameters simultaneously. The experimental results show that BCDDM can not only generate clear and high-quality black hole images from noisy images, but also ensure that the features of the generated images correspond accurately to the initial parameters. We used NRMSE and SSIM to quantify the similarity between GRRT images and BCDDM-generated images with the same physical parameters and found that all reconstructed images reveal high SSIM values. However, the spatial alignment between BCDDM-generated and GRRT images occasionally varies, resulting in some images having high NRMSE, which is likely due to the influence of the initial noise distribution during sampling from pure Gaussian noise. In addition, the model has limited ability to reconstruct images beyond the parameter range of the original dataset, indicating that the current model mainly focuses on the parameter space it has been trained on. However, this limitation can be mitigated by expanding the parameter range of the training dataset.

In addition, our research also indicates that the dataset generated by BCDDM is highly consistent with the training set in terms of statistical properties and physical significance. By mixing generated data with real data for training a regressor, we found that its performance was significantly better than a regressor trained solely on real data. This improvement is attributed to the significant increase in the number and diversity of the training set generated by the data, thereby improving the model's generalization ability.
In the experiment, we used a single RTX 3090 GPU to generate images with a specific radiative flux, achieving an efficient performance with a generation time of just 5.25 seconds per image. Although the simulated dataset used in this study is derived from the RIAF model, the BCDDM method is broadly applicable and can be extended to data from other accretion disk model, with future applications anticipated for a wider range of accretion disk models. In addition, the polarization information in the black hole image can provide more information for analyzing the properties of the accretion flow around the black hole \citep{2021ApJ...910L..12E, 2021ApJ...910L..13E, 2024ApJ...964L..25E, 2024ApJ...964L..26E}. The images input into our model are single-channel and do not consider the polarization information of the images. However, encoding the polarization information into multi-channel images and using neural networks such as the Denoising Diffusion Model to generate polarization images can be further explored in future work.

\section*{Acknowledgments}
This work was supported by the National Natural Science Foundation of China under Grants No. 12374408, 12475051, 12275078, 12035005;  the Natural Science Foundation of Hunan Province under grant No. 2023JJ30384;  the science and technology innovation Program of Hunan Province under grant No. 2024RC1050; and the innovative research group of Hunan Province under Grant No. 2024JJ1006. The authors thank the anonymous referee for their helpful suggestions regarding the manuscript.

\section*{Data and Code Availability}
\textbf{Data Availability: }The datasets used and/or analyzed during the current study are publicly available at \url{https://doi.org/10.5281/zenodo.15354648}.

\textbf{Code Availability: }The code and implementation details for this study are publicly available at \url{https://github.com/LAAAAAAAAA/BCDDM}.

\newcommand{\JournalTitle}{}
\bibliography{main}{}
\bibliographystyle{aasjournal}

\end{document}